\documentclass[prb,preprint]{revtex4-1} 

\usepackage{amsmath}  % needed for \tfrac, \bmatrix, etc.
\usepackage{amsfonts} % needed for bold Greek, Fraktur, and blackboard bold
\usepackage{color}
\usepackage{graphicx}

\begin{document}

\title{Geometric adiabatic angle in anisotropic  oscillators}
\author{Fumika Suzuki}
\affiliation{Theoretical Division, Los Alamos National Laboratory, Los Alamos, New Mexico 87545, USA}
\affiliation{Center for Nonlinear Studies, Los Alamos National Laboratory, Los Alamos, New Mexico 87545, USA}

\author{Nikolai A. Sinitsyn}
\affiliation{Theoretical Division, Los Alamos National Laboratory, Los Alamos, New Mexico 87545, USA}

\begin{abstract} 
We discuss a classical anisotropic oscillator and the Foucault pendulum as examples  illustrating non-conservation of action variables in integrable classical mechanical systems with adiabatically slow evolution. We also emphasize the importance of the mass parameter of a harmonic oscillator, alongside its frequency, in explicitly time-dependent situations.    
\end{abstract}

\maketitle

\section{Introduction}

In this article, we highlight a   misinterpretation that could result from reading physics literature on the conservation of  adiabatic invariants amid slow changes of parameters in classical integrable systems with  multiple ($N>1$) degrees of freedom. These  systems can be separated, by a certain variable transformation, into independent one-dimensional subsystems with coordinates $x_k$ and momenta $p_k$, such that the classical mechanical Hamiltonian is given by
\begin{equation}
H=\sum_{k=1}^N H_k(x_k,p_k).
\label{Hsep}
\end{equation}

For any subsystem with one degree of freedom, there is a canonical transformation to a new momentum, $I_k$, given by
\begin{equation}
    I_k=\frac{1}{2\pi} \oint p_k(x) \,dx_k,
    \label{inv}
\end{equation}
where the integral is along the system's periodic trajectory within $(x_k,p_k)$ subspace. We denote the canonically conjugate to $I_k$ angle variable as  $\theta_k$. The one-dimensional Hamiltonian $H_k$ depends only on $I_k$ but does not depend on $\theta_k$. Therefore, in the action-angle variables, the evolution  is trivial: 
$$
I_k={\rm const}_k, \quad \theta_k(t)=\omega_k(I_k)t,  \quad k=1,\ldots, N,
$$
where 
\begin{equation}
  \omega_k(I_k) \equiv \frac{\partial H_k(I_k)}{\partial I_k}
  \label{freqs}
\end{equation}
are the corresponding frequencies.

Let us now allow some of the parameters of a classical Hamiltonian to be adiabatically-slowly time-dependent, ensuring that if time $t$ is considered as a constant parameter, the Hamiltonian stays separable. A standard result, found in many textbooks on classical mechanics \cite{landau,arnold} is that the action
defined by Eq.~(\ref{inv}) for a one-dimensional motion
turns into an adiabatic invariant, meaning it is conserved up to minor, exponentially small nonadiabatic corrections \cite{note}.

The generalization of this result to  integrable systems with $N>1$ is often deemed trivial. This creates an impression that each of the components $I_k$, defined in Eq.~(\ref{inv}), is  conserved separately. For example, the volume of {\it Mechanics}  by Landau and Lifshitz \cite{landau} addresses  adiabatically time-dependent integrable multi-dimensional systems, and states that $I_k$ are conserved throughout the adiabatic evolution, without acknowledging potential exceptions.  

Here, we note that the trivial arguments that lead to conservation of all $I_k$ apply when $N>1$ only when the  transformation that separates variables in the Hamiltonian of Eq.~(\ref{Hsep}) is time-independent or the frequencies $\omega_k$ are all incommensurate, i.e., their ratios are not rational numbers. When neither of these conditions are satisfied, during adiabatic evolution, $I_k$ can generally  change with time substantially.

Specifically, for commensurate frequencies, the definition of the actions $I_k$ is ambiguous \cite{landau}, and  may also vary over time if a variable transformation that reduces the Hamiltonian to the form of Eq.~(\ref{Hsep}) is time-dependent. This may result in  non-conservation of the individual $I_k$ even at the end of slow cyclic evolution of the  time-dependent parameters,  as for the Berry phase in quantum mechanics. That is, upon the completion of the periodic cycle in space of control parameters, the actions $I_k$, as defined in Eq.~(\ref{inv}), generally do not revert to their initial values but instead gain a geometric contribution that depends on the path chosen for the cyclic parameter evolution. 

We do not claim scientific novelty for this observation. For example, the rotation of the plane of oscillation of the Foucault pendulum can be interpreted as the change of the adiabatic invariant components
$$
I_{x}=\frac{1}{2\pi} \oint P_x \,dX, \quad I_{y}=\frac{1}{2\pi} \oint P_y \,dY,
$$
where coordinates $(X,Y)$ and momenta $(P_x,P_y)$ are defined in a rotating with the Earth frame of a local observer of the pendulum motion. However, the Foucault pendulum is usually described  without references to a more general phenomenon of non-conservation of individual $I_k$ in multi-dimensional integrable systems. For example, in Ref.~\cite{khein}, after switching to cylindrical coordinates, the Foucault rotation angle was interpreted in terms of the Hannay angle \cite{hannay}  in effectively one-dimensional motion. There have also been thorough mathematically rich discussions of the Hannay angle generalization \cite{gen-Hannay} and other geometric effects in classical mechanics \cite{marsden}, which should implicitly contain the physics that we consider here.  This literature, however,  is hard to follow without proper experience with mathematical notation.

Therefore, we believe that the non-conservation of individual $I_k$  in multi-dimensional integrable systems requires a better clarification, in the spirit of a physics textbook, such as the Landau-Lifshitz's volume \cite{landau}.  Our goal is to provide an easy to follow example illustrating this effect, and suggest a more physically intuitive approach to describe it in other classical mechanical systems, such as the Foucault pendulum.

\section{Elementary harmonic oscillator}
Before we introduce our model, let us here recall that
the Hamiltonian of a one-dimensional classical harmonic oscillator contains two parameters: the mass, $m$, and the oscillation frequency, $\omega$:
\begin{equation}
 H_{\rm osc}(x,p) = \frac{p^2}{2m}+\frac{\omega^2 m x^2}{2}, \quad \omega = \sqrt{\frac{k}{m}},
 \label{Hosc-0}
\end{equation}
where $k$ is the spring constant; $x$ and $p$ are the coordinate and the momentum.
For this oscillator, the Hamiltonian in the action-angle variables is particularly simple \cite{landau}: 
\begin{equation}
   H_{\rm osc}(I) = \omega I.
    \label{HamI}
\end{equation}

It is important to note here that the classical Hamiltonian (\ref{Hosc-0}) depends on two independent parameters, $m$ and $\omega$, while the Hamiltonian~(\ref{HamI}) depends explicitly  on only one parameter $\omega$. There is no contradiction here because the equations  defining the transformation $(x,p)\rightarrow (\theta, I)$ depend on $m$:
\begin{equation}
    x(\theta, I)=\sqrt{\frac{2I}{m\omega}}\sin \theta, \quad p(\theta,I)=\sqrt{2I\omega m}\cos \theta.
    \label{xpI}
\end{equation}
Hence, implicitly, the Hamiltonian (\ref{HamI}) depends on $m$ through the dependence of $I$ on it in the original variables $(x,p)$. In what follows, we show that the mass parameter $m$ is not redundant when considering explicitly time-dependent evolution.

\section{Harmonic oscillator with anisotropic mass}

%%%%%%%%%%%%%%%%%%%%%%%%%%%%%%%%
\begin{figure}
\centering
{%
\includegraphics[clip,width=0.5\columnwidth]{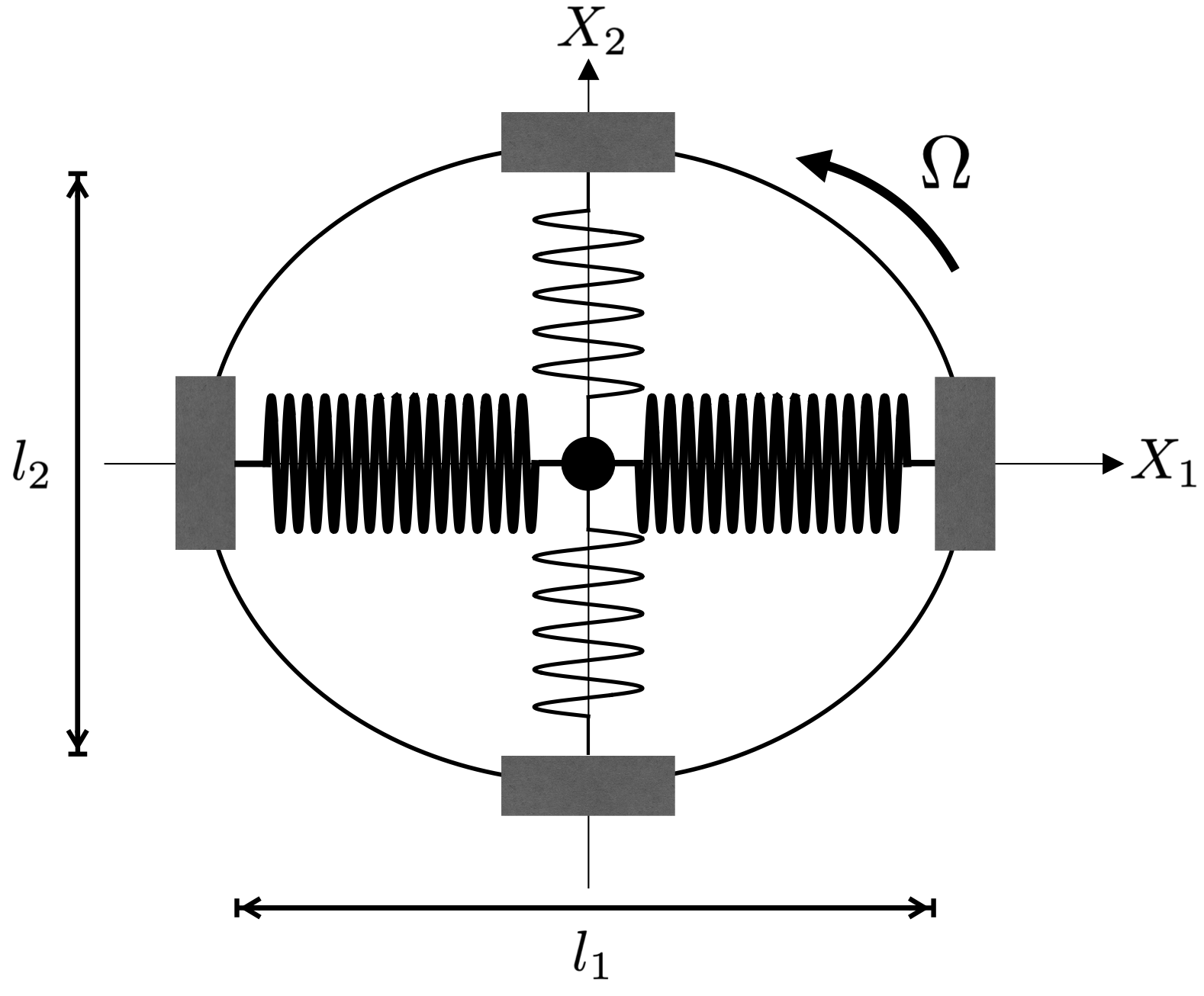} 
}
\caption{ Two non-ideal spring pairs attached to the walls. The walls are rigidly connected by a hoop, so the  distances between them are fixed. The springs are also joined rigidly at their crossing point (black circle), possibly with an extra mass attached to it. Small perturbations lead to harmonic oscillations of the joint  near its equilibrium position; $X_1$ and $X_2$ are the displacements along the horizontal and the vertical springs, respectively. The entire structure slowly rotates around the equilibrium point. The horizontal and vertical springs are made of different materials but the distances between the opposite walls, $l_1$ and $l_2$, are chosen such that the  oscillations of $X_1$ and $X_2$ have the same frequency, $\omega$. Energy dissipation and anharmonicity effects are disregarded.  
}
\label{spring-cross}
\end{figure}
%%%%%%%%%%%%%%%%%%%%%%%%%%%%%%%%%%%%%
Our model is a two-dimensional harmonic oscillator with an anisotropic mass. The mass anisotropy is common in physics, e.g., electrons in crystals are described as quasi-particles with an effective mass that is different for forces acting along different directions.  Usually, an effective  mass is introduced in equations of motion of a collective excitation of many particles.  An example is dynamics of the length of a  non-ideal spring with a finite intrinsic weight.

A simple classical system that may be described with anisotropic mass is shown in Fig.~\ref{spring-cross}.  This system is composed of four massive springs joined at a central point.  The vertical springs are identical. The horizontal springs are identical to each other, but not to the vertical springs. For example, the vertical and horizontal springs in Fig.~\ref{spring-cross} can be made of different materials. For sufficiently slow disturbances, the motion of a massive spring can be well approximated by an ``effective mass" approximation, where the massive spring is replaced by an ideal massless spring with an effective mass at the end \cite{AJP1984,RodriguezGesnouin2007}. We also assume that the springs are firmly joined  at their crossing point. This point may not have any significant mass on its own.

Let us denote the deviation of the disturbed spring joint from its equilibrium position along the line of the horizontal springs, shown in Fig.~\ref{spring-cross}, as $X_1$, and the deviation along the line of the vertical springs  as $X_2$. Forces that behave as higher than linear powers of $X_1$ or $X_2$ may be present, e.g., a displacement of the horizontal spring in the vertical direction generally introduces a restoring force proportional to the cube of the displacement \cite{Hinrichsen-23}. However, such forces become negligible, in comparison to the linear forces, for sufficiently small spring oscillation amplitudes. Therefore, such oscillations are harmonic; the linear-in-displacement forces are aligned with the spring axes.  Thus, we will disregard both the nonlinear forces and energy dissipation.

If the walls in Fig.~\ref{spring-cross} are fixed, the effective equations of motion for the joint point near its equilibrium position are given by 
\begin{equation}
  m_1 \ddot{X}_1 = -k_1 X_1, \quad m_2 \ddot{X}_2 = -k_2 X_2.
    \label{fix-em}
\end{equation}
Since the effective mass-like parameters, $m_{1,2}$, and the effective spring constants, $k_{1,2}$,  depend  on the materials from which the springs are made, generally, $m_1\ne m_2$, and  Eq.~(\ref{fix-em}) effectively describes a two-dimensional harmonic oscillator with an anisotropic mass.   Here, we do not discuss details for derivation of the linear forces leading to Eq.~(\ref{fix-em}). Anisotropic harmonic oscillations follow just from the symmetry of the problem and the fact that, near the equilibrium point $(X_1,X_2)=(0,0)$, the potential energy as a function of $X_1$ and $X_2$ must be a quadratic function of $X_1$ and $X_2$ up to disregarded nonlinear terms.

We will also assume that the geometric parameters of the springs are tuned so that the intrinsic oscillation frequencies along both spring axes are equal:
\begin{equation}
  \sqrt{\frac{k_1}{m_1}}= \sqrt{\frac{k_2}{m_2}} \equiv \omega.
    \label{freq1}
\end{equation}
%%%%%%%%%%%%%%%%%%%%%%%%%%%%%%%%%%%%%
Thus, if the walls in Fig.~\ref{spring-cross} are fixed, the spring system Hamiltonian is given by 
\begin{equation}
H=\frac{P_1^2}{2m_1}+\frac{P_2^2}{2m_2} +\frac{\omega^2 m_1 X_1^2}{2} +\frac{\omega^2m_2 X_2^2}{2},
    \label{Ham-spr}
\end{equation}
where $P_{1,2}$ are canonically conjugate to $X_{1,2}$ momenta. 
 We assume here that $\omega$ is the lowest excitation frequency in the spring system, so for adiabatic evolution of parameters and a slowly introduced initial perturbation the higher frequency modes related to emergence of wave-like dynamics \cite{Dooling} are strongly suppressed, and we will disregard them. 
Note also that in terms of the action variables the Hamiltonian (\ref{Ham-spr}) is particularly symmetric
\begin{equation}
H=\omega(I_1+I_2 ).
    \label{sym-Ham}
\end{equation}

\section{Oscillations in rotating frame}

Consider the situation in which the walls in Fig.~\ref{spring-cross} are rigidly connected by a hoop, and the entire structure  performs a slow rotation around the equilibrium point of the spring joint.
Relative distances between the walls are kept constant. 

Let $(x,y)$ be a fixed relative to distant stars, i.e., time-independent, frame of reference. The coordinates $(X_1,X_2)$ and $(x,y)$ are then related by
\begin{equation}
x=X_1\cos\varphi(t) -X_2\sin\varphi(t), \quad y = X_1\sin\varphi(t) +X_2 \cos \varphi(t),
\label{rotate-x}
\end{equation}
where $\varphi(t)$ is the rotation angle of the entire structure around the joint equilibrium point.
We assume that the rotation of the entire structure is very slow, e.g., 
$$
\varphi(t) =\Omega t, \quad \Omega \ll \omega.
$$ 

Let us also introduce the  momenta $(p_x,p_y)$ that are canonically conjugate to $(x,y)$. During the frame rotation, the momentum vector is transformed the same way as the vector of coordinates:
\begin{equation}
p_x=P_1\cos\varphi(t) -P_2\sin\varphi(t), \quad p_y = P_1\sin\varphi(t) +P_2 \cos \varphi(t).
\label{rotate-p}
\end{equation}
The inverse transformation is given by:
\begin{eqnarray}
\label{rotate-XX}
 X_1&=&x\cos\varphi(t) +y\sin\varphi(t), \quad X_2 = -x\sin\varphi(t) +y \cos \varphi(t),\\
\nonumber P_1&=&p_x\cos\varphi(t) +p_y\sin\varphi(t), \quad P_2 = -p_x\sin\varphi(t) +p_y \cos \varphi(t).
\label{rotate-PP}
\end{eqnarray}
Thus, if we treat time $t$ as a parameter, the Hamiltonian is written in terms of the variables $(x,p_x)$ and $(y,p_y)$   by substituting Eq.~(\ref{rotate-XX}) into Eq.~(\ref{Ham-spr}).  This Hamiltonian  does not have the form of Eq.~(\ref{Hsep}). However, it can be transformed to the form of Eq.~(\ref{Hsep}) by switching the variables back using Eqs.~(\ref{rotate-x}) and (\ref{rotate-p}). 

In the fixed frame, the variable pairs $(x,p_x)$ and $(y,p_y)$ no longer have independent dynamics,  e.g., the Hamiltonian, for $m_1\ne m_2$,  contains cross-interaction terms $\propto xy$ and $\propto p_xp_y$. The origin of this interaction can be traced to the effect of the constraint introduced by the rigid joint of the springs. Thus, the symmetry of the  Hamiltonian, as expressed in Eq.~(\ref{sym-Ham}), can be misleading when, for some reason, this system must be considered in a time-dependent frame. 

\section{Action-angle variables in rotating frame}
\label{aa-sec}

Let us interpret the relations (\ref{rotate-x}) and (\ref{rotate-p}) as a canonical variable transformation that brings the Hamiltonian $H(x,y;p_x,p_y)$ of the rotating structure back into the form (\ref{Ham-spr}), in which the variables $(X_1,P_1)$ and $(X_2,P_2)$ are uncoupled.
When $\varphi(t)$ is not a constant parameter, the Hamiltonian acquires a non-adiabatic correction, which can be inferred from the invariance of the action 
\begin{equation}
S=\int \{p_x\, dx +p_y\,dy -H(x,y;p_x,p_y)\, dt \}  
\label{action}
\end{equation}
under canonical transformations.
Substituting (\ref{rotate-x}) and (\ref{rotate-p}) into (\ref{action}), we find
\begin{equation}
S=\int \{P_1\, dX_1 +P_2\,dX_2 -H' dt \},   
\label{action-1}
\end{equation}
where, in terms of $(X_1,P_1)$ and $(X_2,P_2)$,  
\begin{equation}
 H'=H+(P_1X_2-P_2X_1)\dot{\varphi}.
    \label{hprime0}
\end{equation}
The last term in Eq.~(\ref{hprime0}) is responsible for fictitious Coriolis and centrifugal forces. By working only with canonical coordinates and momenta,
we will not separate individual effects of these forces. 
We only mention that the centrifugal force, being quadratic in the rotation frequency, $\dot{\varphi}$, plays a negligible role for the adiabatic rotation considered here, so the following geometric effects can be considered as the consequences of the Coriolis force.  

Switching to the action-angle variables,
\begin{equation}
    X_{1,2}=\sqrt{\frac{2I_{1,2}}{m_{1,2}\omega}}\sin \theta_{1,2}, \quad P_{1,2}=\sqrt{2I_{1,2}\omega m_{1,2}}\cos \theta_{1,2},
\label{actI}
\end{equation}
 we find

\begin{equation}
H'=\omega(I_1+I_2)+2\sqrt{I_1I_2}
\left(
\sqrt{\frac{m_1}{m_2}}\cos \theta_1 \sin \theta_2-\sqrt{\frac{m_2}{m_1}} \cos \theta_2 \sin \theta_1  \right)\dot{\varphi}.
    \label{hprime}
\end{equation}
Thus, in the action-angle variables the nonadiabatic, generally time-dependent via $\varphi(t)$, correction to the Hamiltonian depends on the difference between the masses $m_1$ and $m_2$.

The equations of motion are given by 
\begin{equation}
\dot{\theta}_{1,2}=\frac{\partial H'}{\partial I_{1,2}}, \quad \dot{I}_{1,2}=-\frac{\partial H'}{\partial \theta_{1,2}}.
\label{eq-motion}
\end{equation}
In these equations, it is convenient to change variables to 
\begin{equation}
\theta=\frac{\theta_1+\theta_2}{2}, \quad \theta_{-} = \theta_1- \theta_{2}, \quad I=I_1+I_2, \quad  I_{-}=I_1-I_2.
    \label{thet-pm}
\end{equation}
Then, we find that $\theta$ is the fast variable that increases with time as
\begin{equation}
    \theta \approx \omega t.
    \label{thet-plus0}
\end{equation}
We also introduce a parameter that characterizes the mass asymmetry:

\begin{equation} 
\mu\equiv \frac{m_1+m_2}{2\sqrt{m_1m_2}}.
    \label{mu-def}
\end{equation}

\section{The motion averaged over fast oscillations}
\label{average-section}

We substitute Eqs.~(\ref{thet-pm}) and (\ref{mu-def}) into  equations of motion (\ref{eq-motion}) and average the result over the period of fast oscillations by disregarding  all terms $\propto \cos(\theta)$ and $\propto \sin (\theta)$. The final result is 
\begin{equation}
\frac{dI}{d\varphi} =0.
\label{Icons0}
\end{equation}
\begin{eqnarray}
\label{delI}
\frac{d I_{-}}{d\varphi}&=&2\mu\sqrt{I^2-I^2_{-}}\cos \theta_{-},\\
\label{del-theta}
\frac{d\theta_{-}}{d\varphi}&=&2\mu\frac{I_{-}}{\sqrt{I^2-I_{-}^2}}\sin \theta_{-}.
\end{eqnarray}

Equation~(\ref{Icons0}) shows that  $I=I_1+I_2$ is the true adiabatic invariant, which is conserved after averaging over the fast oscillations.   The evolution of $\theta_{-}$ and $I_{-}$, in Eqs.~(\ref{delI}) and (\ref{del-theta}),  does not depend on $t$ explicitly. The changes of  $\theta_{-}$ and $I_{-}$ are entirely determined by the changes of the control parameter $\varphi$. In this regard, these variables are akin to the non-Abelian geometric phase in quantum mechanics. In our case, however, Eqs.~(\ref{delI}) and (\ref{del-theta}) possess an integral of motion that results in an effectively Abelian geometric phase, as we now demonstrate.

Equations~(\ref{delI}) and (\ref{del-theta}) are  the canonical equations of motion for the effective Hamiltonian
\begin{equation}
    {\cal H}=-2\mu \sqrt{I^2-I^2_{-}}\sin \theta_{-},
\label{eff-H}
\end{equation}
in which  $\theta_{-}$ is the coordinate and $I_{-}$ is the momentum. The function
 ${\cal H}$ in Eq.~(\ref{eff-H}) does not depend on $\varphi$ explicitly, so the corresponding energy is conserved. Let us choose initial conditions such that when $\varphi=0$ we have 
$$
I_1\rightarrow I, \quad I_2\rightarrow 0.
$$
This corresponds to $I_{-}\rightarrow I$, and hence to ${\cal H}=0$. By changing $\varphi$, the values of $I_{-}$ deviate from $I$, so  energy conservation requires that for such initial conditions we also have 
\begin{equation}
\theta_{-}=\pi n,
\label{theta-minus}
\end{equation}
where $n$ is an integer. For one of these $\theta_{-}$, we have
\begin{equation}
\label{delI-2}
\frac{d I_{-}}{d\varphi}=2\mu\sqrt{I^2-I^2_{-}} \cos \theta_{-}, \quad I_{-}(0)=I.
\end{equation}
We search for the solution of Eq.~(\ref{delI-2}) in the form 
$$
I_{-}=I\cos (\psi(\varphi)), \quad \psi(0)=0,
$$
where $\psi (\varphi)$ is a new angle variable.
Note that at $I_{-}=-I$ or $I_{-}=I$, the phase $\theta_{-}$ is not well defined because one of the actions $I_{1,2}$ is zero. Hence, by passing through such values of $I_{-}$, the solution can switch between the discrete values in Eq.~(\ref{theta-minus}). This phase shift is inferred from the requirement that the variable $\psi$ changes with time continuously. Thus, each time $\psi$ crosses $0$ or $\pi$, we should assume that $\theta_{-}$ changes by $\pi$. 
This leads to a continuous evolution of $\psi$:
\begin{equation}
 \frac{d \psi}{d\varphi}=-2\mu, 
 \label{connection1}
\end{equation}
with a trivial solution
\begin{equation}
\psi(\varphi)=-2\mu \varphi,
    \label{sol1}
\end{equation}
and 
$\theta_{-}=\pi (1+[\psi/\pi])$, where $[...]$ is the integer part of the expression in the brackets.

\section{The geometric shift at the end of the cycle}
 Equation~(\ref{connection1}) can be interpreted as the relation 
$$
d\psi = Ad\varphi,
$$
where $A=-2\mu$ is an Abelian connection. As $\varphi$ changes from $0$ to $2\pi$, the walls in the structure return to their initial places, so the change of the actions can be defined unambiguously. We find
\begin{equation}
\Delta I\equiv I_{-}(2\pi) = I\cos\left( \oint A\, d\varphi \right) = I\cos\left(2\pi \frac{m_1+m_2}{\sqrt{m_1m_2}}\right).
\label{delI-fin}
\end{equation}
From Eq.~(\ref{delI-fin}), the individual actions after completion of the cycle are given by
\begin{eqnarray}
\label{delI1-fin}
\nonumber I_1^{\rm fin}&=& \frac{I+\Delta I}{2}=I\cos^2(2\pi \mu), \\
 I_2^{\rm fin}&=& \frac{I-\Delta I}{2}=I\sin^2(2\pi \mu).
\end{eqnarray}

The case with $m_1=m_2$ corresponds to $\mu=1$. As expected, without the mass anisotropy, e.g., for the springs made of the same material, the actions do not change.  Equation~(\ref{actI}) provides a simple approach to test Eqs.~(\ref{delI1-fin}). Let us start with oscillations  along the $x$-axis, which correspond to $X_2=0$ in the rotating frame.   After the structure is slowly rotated by an angle $2\pi$, the oscillations end up to be along a line rotated by an angle, $\alpha$, such that
\begin{equation}
\tan \alpha \equiv \frac{X_{2}}{X_{1}} =\pm \sqrt{\frac{m_1I_{2}^{\rm fin}}{m_2I_{1}^{\rm fin}}},
\label{xx-meas}
\end{equation}
where $\pm$ corresponds to, respectively, odd and even number of passages through  zero value of one of the actions $I_{1}$ and $I_2$. 

To obtain $\alpha$ numerically, we chose $X_{1,2}$ in Eq.~(\ref{xx-meas}) to be the spacial turning points of the final trajectory. Thus, by measuring $\tan \alpha$, one can determine the ratio $\sqrt{I_{2}^{\rm fin}/I_{1}^{\rm fin}}$, and then compare the result with the prediction of Eq.~(\ref{delI1-fin}).

\begin{figure}
\centering
{%
\includegraphics[clip,width=0.5\columnwidth]{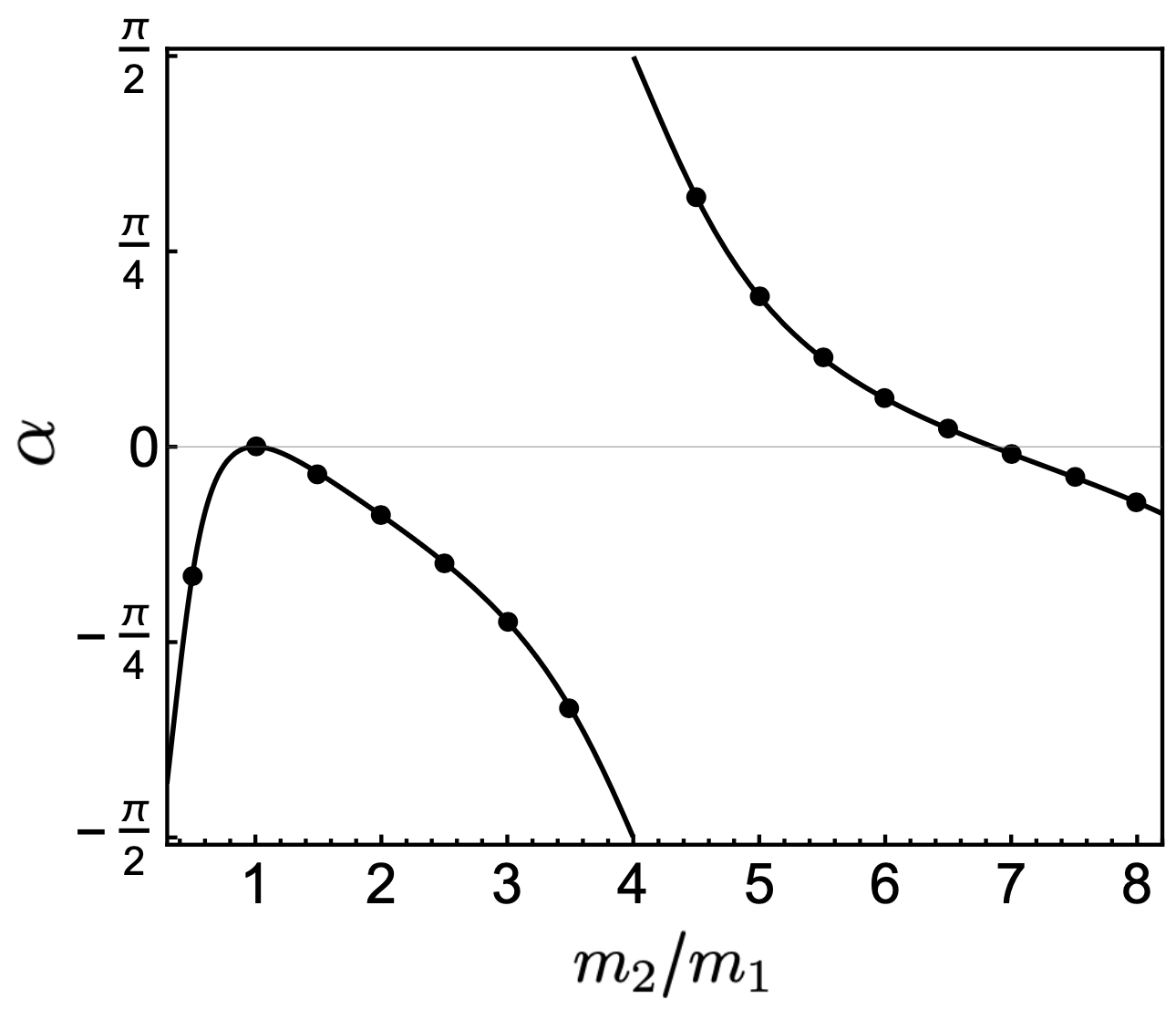} 
}
\caption{Numerical test of Eq.~(\ref{xx-meas}): The geometric rotation angle $\alpha$ as the function of $m_2/m_1$. The  sign change of $\alpha$ should be interpreted as a mark for an additional $\pi$-shift in $\theta_{-}$ after one of $I_{1,2}$ touches zero value.
The solid curves represent the analytical prediction. The dots are obtained from the numerical solutions of the  evolution equations with the time-dependent Hamiltonian, considered in the fixed frame with $\varphi(t)=\pi (1+\tanh\Omega t)$, where $t\in (-T,T)$ and $T\gg 1/\Omega$. Here, we used that, according to Eq.~(\ref{actI}),  $\pm \sqrt{I_2^{\rm fin}/I_1^{\rm fin}}=\sqrt{m_2}X_2/\sqrt{m_1}X_1$, which relates $I_{1,2}^{\rm fin}$ to the values of $X_{1,2}$ at the end of the structure rotation.
}
\label{fig2}
\end{figure}

In Fig.~\ref{fig2}, we confirm the predictions in Eq.~(\ref{xx-meas}) with the results of our direct numerical simulations of the Hamiltonian equations of motion  after a full rotation of the structure over an angle $2\pi$. A  rotation angle, $\varphi(t)=\pi(1+\tanh \Omega t)$, where $t$ was changing from large negative to large positive values, was chosen to switch the rotation on and off smoothly in order to avoid $\sim \Omega/\omega$ nonadiabatic boundary effects. 

\section{Precession of elliptic trajectory and its area conservation}

\begin{figure}
\centering
{%
\includegraphics[clip,width=0.5\columnwidth]{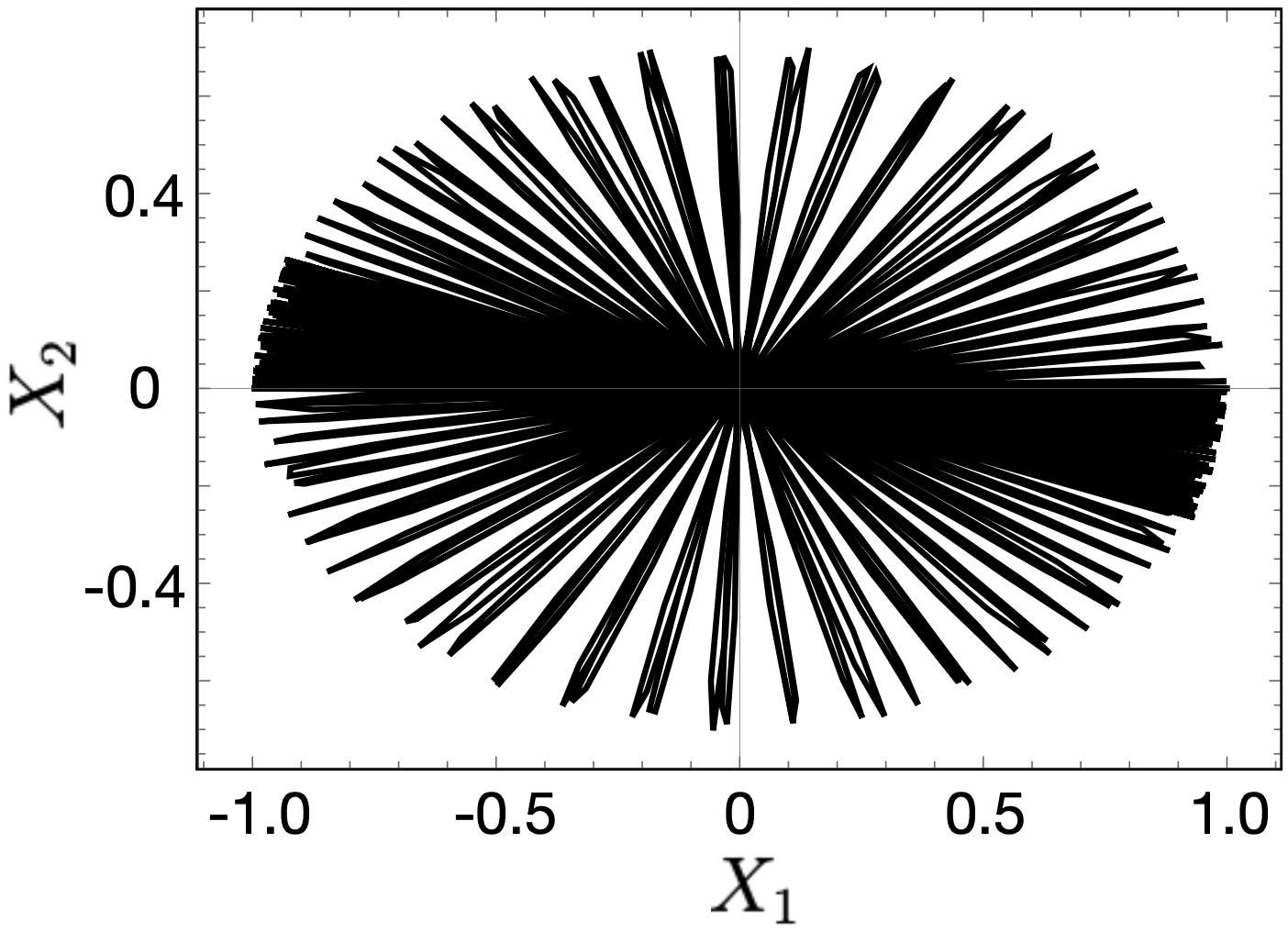} 
}
\caption{Numerical test of  the area conservation for periodic trajectories in real space. Here, the rotation of the structure is performed with angular velocity $\Omega=0.01\omega$, where $\omega=1$ at  $m_1=1,\, m_2=2$ and the initial conditions $x=X_1(t=-1000)=1,\, y=X_2(t=-1000)=0$, and $p_x=P_1(t=-1000)=p_y=P_2(t=-1000)=0$; this corresponds an almost zero area inside the periodic  trajectories (up to small nonadiabatc corrections $\sim \Omega/\omega$).
This property is conserved during the   rotation of the structure. The  evolution trajectory is obtained initially in the fixed frame  for $\varphi(t)=\pi (1+\tanh\Omega t)$, and then the trajectory in variables ($X_1,X_2$) is found using Eq.~(\ref{rotate-XX}). Note that while $\varphi \in (0,2\pi]$, the line of oscillations in the rotating frame rotates by an angle different from $2\pi$. This difference is the rotation of the trajectory that would be observed from the fixed frame. 
%(upper panel), $P_1(t=-1000)=0, P_2(t=-1000)=0.5$ (lower panel).
}
\label{fig3}
\end{figure}

The simplicity of the solution in Eq.~(\ref{delI-fin}) follows from the conservation of the pseudo-energy ${\cal H}$, i.e.,  the right hand side in Eq.~(\ref{eff-H}), during the adiabatic evolution. This means that ${\cal H}$ is an additional adiabatic invariant. It has a simple physical interpretation. From the relation of the variables in Eq.~(\ref{xpI}), the trajectory of the joint point in the real space is given by 
\begin{equation}
 X_{1}(t)=\sqrt{\frac{2I_{1}}{m_1\omega}}\sin(\omega t+\theta_{-}/2), \quad X_2(t) =\sqrt{\frac{2I_{2}}{m_2\omega}}\sin(\omega t-\theta_{-}/2).
\label{real-tr}
\end{equation}

Equation~(\ref{real-tr}) shows that at fixed $I_{1,2}$, the real space trajectory of our oscillator would be an ellipse, with amplitudes along the frame axes $\propto \sqrt{I_{1,2}}$. Therefore, as the relative values of $I_{1,2}$ slowly change, these amplitudes change too, which should be interpreted as the rotation of the orientation of the main axes of this elliptic trajectory in real space. This rotation is  shown in numerically generated Fig.~\ref{fig3} for the initial conditions such that the ellipse almost degenerates into a thin line.

The area inside the elliptic trajectory is 
$$
S_{12}=\oint X_1\, dX_2 \propto \sqrt{I_1I_2} \sin \theta_{-} \propto \sqrt{I^2-I_{-}^2} \sin \theta_{-},  
$$
which means that, up to a constant factor, ${\cal H}$ is $S_{12}$, so  $S_{12}$ is an adiabatic invariant.  We confirm this property in Fig.~\ref{fig3}, which demonstrates that the periodic trajectory with initially zero area in real space remains looking as a straight piece of a line, despite the adiabatic rotation that this line makes. 

\section{3D-generalizations and Foucault pendulum }
These simplifications do not  appear in the higher-dimensional version of our model generally. For example, in three dimensions, our structure has  different non-ideal springs along three orthogonal axes, whose directions vary in the fixed  $xyz$-frame. 
The rotation of the structure around the equilibrium of the joint point of the three springs is then generally parametrized by two angles in spherical coordinates: $(\varphi, \vartheta)$. For all identical frequencies, only the combination $I=I_1+I_2+I_3$ is then the true adiabatic invariant, and the corresponding fast angle variable is $(\theta_1+\theta_2+\theta_2)/3$. However, there are now four instead of two coupled equations for the  independent   angle-action variables, such as $I_1-I_2$, $I_1-I_3$, $\theta_1-\theta_2$, and $\theta_1-\theta_3$. Moreover, these equations now depend non-trivially on two control parameters $\varphi$ and $\vartheta$. Therefore, even after averaging over fast oscillations, the result of the time evolution must be obtained by considering the  equations that may not generally have an analytical solution. 

Such  changes of the  actions  are    geometric in the sense that they  depend on the geometry of the control protocol but not depend on the rate of the passage over it, as long as this passage is sufficiently slow. They are, therefore, the direct classical mechanical counterpart of the non-Abelian geometric phase in quantum mechanics. 

\begin{figure}
\centering
{%
\includegraphics[clip,width=0.4\columnwidth]{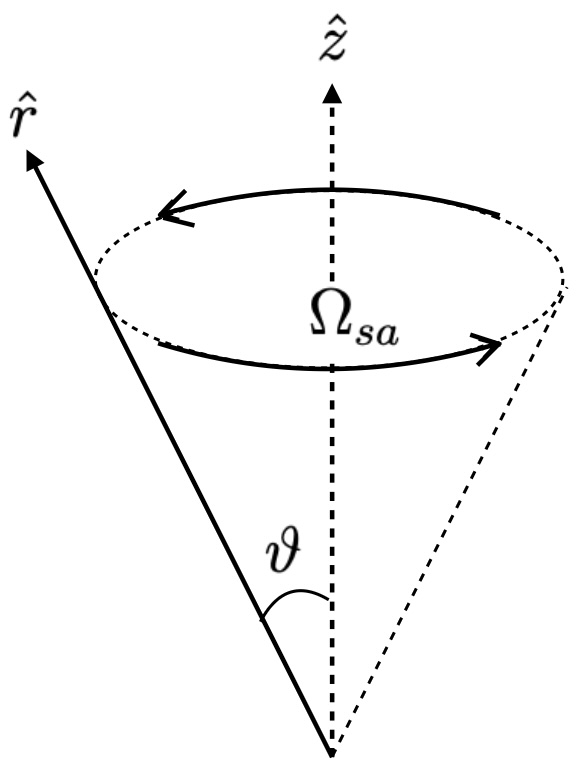} 
}
\caption{The unit vector $\hat{r}$, pointing along the frequency anisotropy axis of the Hamiltonian (\ref{F-ham}),  is slowly rotating around the vector $\hat{z}$, pointing along the $z$-axis in the fixed frame. The angle $\vartheta$ does not change with time, so by completion of the circle, the vector $\hat{r}$ describes an area on the unit sphere that is viewed from the origin as subtending a solid angle $\Omega_{sa}=2\pi(1- \cos \vartheta)$.   
}
\label{fou}
\end{figure}

The  Foucault pendulum is described by essentially the same equations as a 3-dimensional version of our structure, in which only two  frequencies are degenerate, while there is no mass anisotropy. Namely, consider an oscillator with the Hamiltonian 

\begin{equation}
H=\frac{P_x^2+P_y^2+P_z^2}{2m} + \frac{m\omega_0^2(X^2+Y^2)}{2}+ \frac{m\omega^2 Z^2}{2},
\label{F-ham}
\end{equation}
where we require that $\omega \ne \omega_0$ in order to guarantee that there is no energy transfer between oscillations in the plane $(X,Y)$ and along the $Z$-axis during the adiabatic evolution.
The Hamiltonian (\ref{F-ham}) shares two properties with the Foucault pendulum: it has a frequency degeneracy for oscillations in plane $(X,Y)$, and an anisotropy axis, $Z$, that performs a rotational motion. One can think about the Foucault pendulum as the limit of $\omega \rightarrow \infty$ in Eq.~(\ref{F-ham}), so that the amplitude of harmonic oscillations along the anisotropy axis is strongly suppressed.  

We assume that the initial conditions have $Z=0$, so the oscillations have a substantial amplitude only within the $XY$-plane.
Let us now assume that the structure performs a rotation that affects directions of all three axes when viewed from the fixed frame $xyz$ with the origin at the spring joint equilibrium point. We split derivation of the geometric rotation angle  for this oscillator in five steps:

\paragraph*{\bf 1. Find the transformation between variables in the fixed and rotating frames.} Let $\hat{x}$, $\hat{y}$, and $\hat{z}$ represent unit vectors along the axes in a fixed coordinate system. We assume that our 3-dimensional structure rotates around the equilibrium joint point, so that the unit vector, $\hat{ r}$, along the $Z$-axis, has components in the fixed frame given by 
\begin{equation}
\hat{ r}=\left(\begin{array}{c}
\sin \vartheta \cos \varphi \\
\sin \vartheta \sin \varphi\\
\cos \vartheta
\end{array}
\right).
    \label{hatr}
\end{equation}
 We will also assume that this direction changes periodically, as shown in Fig.~\ref{fou}. For simplicity, we consider that $\vartheta={\rm const}$, and $\varphi$ changes in the interval $\varphi \in [0,2\pi]$, so that the vector $\hat{r}$ sweeps a solid angle on the unit sphere equal to 
 $$
 \Omega_{sa}=2\pi(1- \cos \vartheta).
 $$

To separate variables in the Hamiltonian, due to the rotational symmetry of interactions around $\hat{r}$, the axes in plane transverse to $\hat{r}$ can be chosen arbitrarily, with the only requirements that corresponding unit vectors along them, $\hat{ \varphi}$ and $\hat{ \vartheta}$, must be mutually orthogonal with each other and $\hat{r}$, and varying periodically with $\varphi$. Thus, we choose them to be
\begin{equation}
  \hat{ \vartheta}=\left(\begin{array}{c}
\cos \vartheta \cos \varphi \\
\cos \vartheta \sin \varphi\\
-\sin \vartheta
\end{array}
\right), \quad 
\hat{\varphi}=\left(\begin{array}{c}
- \sin \varphi \\
 \cos \varphi\\
0
\end{array}
\right), 
    \label{hatangles}
\end{equation}
so that
$$
\hat{\theta} \times \hat{\varphi} = \hat{r}, \quad \hat{r} \times \hat{\vartheta} = \hat{\varphi}, \quad \hat{\varphi} \times \hat{r} = \hat{\vartheta}.
$$

Note also that at $\phi=\theta=0$, the vectors $\hat{\vartheta}$, $\hat{\varphi}$, $\hat{r}$ coincide with, respectively, $\hat{x}$, $\hat{y}$, and $\hat{z}$. In the frame with axes along these vectors, the  coordinate and effective momentum vectors of the joint point are given by 
\begin{eqnarray}
\nonumber{\bf r} &=&X\hat{\vartheta}+Y\hat{\varphi}+Z\hat{r},\\ 
\label{rp-vec}
{\bf P} &=&P_x\hat{\vartheta}+P_y\hat{\varphi}+P_z\hat{r},
\end{eqnarray}
and the Hamiltonian for the motion of the spring joint has the variable-separated form of Eq.~(\ref{F-ham}). 

Let, in the fixed frame, the same coordinate and momentum vectors have components, respectively, ${\bf r}_0=(x,y,z)$ and ${\bf p}_0=(p_x,p_y,p_z)$. From Eqs.~(\ref{hatr}), (\ref{hatangles}), and (\ref{rp-vec}), we find then the  transformation that separates the variables of the rotated structure Hamiltonian:
\begin{equation}
\left(\begin{array}{c}
x\\
y\\
z
\end{array}
\right)=R(\vartheta,\varphi) \left(\begin{array}{c}
X\\
Y\\
Z
\end{array}
\right), \quad 
\left(\begin{array}{c}
p_x\\
p_y\\
p_z
\end{array}
\right)=R(\vartheta,\varphi) \left(\begin{array}{c}
P_x\\
P_y\\
P_z
\end{array}
\right),
\label{rot-tr}
\end{equation}
where $R(\vartheta, \varphi)$ is a rotation matrix given explicitly by
\begin{equation}
R(\vartheta, \varphi) =
\begin{bmatrix}
\cos \vartheta \cos \varphi  & -\sin \varphi & \sin \vartheta \cos \varphi \\
 \cos \vartheta \sin \varphi &  \cos \varphi & \sin \vartheta \sin \varphi \\
 -\sin \vartheta& 0  & \cos \vartheta
\end{bmatrix}.
\label{Rmt}
\end{equation}
An easy way to obtain this matrix is to note that its columns coincide with  the vectors  $\hat{\vartheta}$, $\hat{\varphi}$, and $\hat{r}$. This matrix also has a property $R^{-1}=R^T$, which is needed to conserve a vector length under rotation.

\paragraph*{\bf 2. Find the nonadiabatic correction to the Hamiltonian written in the rotating frame.} When the angle $\varphi$ slowly varies with time, the Hamiltonian acquires a nonadiabatic correction
$$
{\cal H}= H({\bf r}_0({\bf r},{\bf P}),{\bf p}_0({\bf r},{\bf P}))+\delta H ({\bf r},{\bf P}),
$$
where $H({\bf r}_0({\bf r},{\bf P}),{\bf p}_0({\bf r},{\bf P}))$ has the form of Eq.~(\ref{F-ham}), and  $\delta H$ is found from the invariance of the action functional: 
$$
S=\int \{{\bf p}_0 \, d{\bf r}_0-H({\bf r}_0,{\bf p}_0)\,dt \},
$$
leading to
\begin{eqnarray}
\nonumber \delta H ({\bf r},{\bf P})&=&-{\bf p}_0({\bf r},{\bf P})\cdot\partial_t{\bf r}_0({\bf r},{\bf P}) = -{\bf P} \cdot R^{T} \frac{dR}{d\varphi} {\bf r} \dot{\varphi}\\
\label{Fhprime}
&=&\{\cos \vartheta (P_xY-P_yX) +\sin \vartheta (P_zY-P_yZ)\}\dot{\varphi}.
\end{eqnarray}

\paragraph*{\bf 3. Switch to the action-angle variables in the rotating frame.} Let the corresponding action-angle varibles be $(I_{\vartheta},I_{\varphi},I_{r})$ and $(\theta_{\vartheta},\theta_{\varphi},\theta_{r})$. In our case, we should substitute
\begin{equation}
X=\sqrt{2I_{\vartheta}/(m\omega_0)} \sin \theta_{\vartheta}, \quad Y=\sqrt{2I_{\varphi}/(m\omega_0)} \sin \theta_{\varphi}, \quad P_z=\sqrt{2I_{r}m\omega} \cos \theta_{r},
\label{var-chage}
\end{equation}
e.t.c., to the Hamiltonian, which then has the form 
\begin{eqnarray}
\nonumber \mathcal{H}&=&\omega_0(I_{\varphi}+I_{\vartheta})+\omega I_{r}-2\cos \vartheta\sqrt{I_{\varphi}I_{\vartheta} }\sin(\theta_{\vartheta}-\theta_{\varphi})\dot{\varphi} \\
\label{calH}
 &+&2\sin \vartheta \sqrt{I_{\vartheta}I_{r}} \dot{\varphi}\left[\sqrt{\frac{\omega}{\omega_0}} \cos \theta_{r}\sin \theta_{\varphi}-\sqrt{\frac{\omega_0}{\omega}} \sin \theta_{r}\cos \theta_{\varphi} \right].
\end{eqnarray}

\paragraph*{\bf 4. Neglect quickly oscillating with zero mean terms in the equations of motion.}
In these equations, the last term, $\propto\sin \vartheta$, in Eq.~(\ref{calH}) leads to quickly oscillating, with zero mean value, contributions because
$$
\theta_r  \approx \omega t,\quad 
\theta_{\varphi} \approx \omega_0t, \quad \theta_{\vartheta} \approx \omega_0t.
$$
Therefore, for $\omega \ne \omega_0$ the terms $\propto \cos \theta_r \cos \theta_{\varphi}$ and $\propto \sin \theta_r \sin \theta_{\varphi}$, e.t.c., average to zero and can be discarded in the adiabatic limit. The variables $I_{r}$ and $\theta_r$ after this decouple, leading to an adiabatic invariant $I_{r}={\rm const}$.

The other nonadiabatic correction, which is $\propto \cos \vartheta$ in Eq.~(\ref{calH}), leads to essentially the same equations as in Section~\ref{average-section}. By introducing $I_{+}=I_{\varphi}+I_{\vartheta}$, $I_{-}=I_{\vartheta}-I_{\varphi}$, and $\theta_+=(\theta_{\varphi}+\theta_{\vartheta})/2$, $\theta_{-}=\theta_{\vartheta}-\theta_{\varphi}$, we find

\begin{equation}
\frac{dI_+}{d\varphi}=0, \quad \frac{d{I}_{-}}{d\varphi}=2\cos \vartheta \sqrt{I^2-I_{-}^2}  \cos \theta_{-}, \quad \frac{d{\theta}_{-}}{d\varphi}= 2\cos \vartheta \frac{I_{-}}{\sqrt{I^2-I_{-}^2}} \sin \theta_{-}.
\label{FIm}
\end{equation}

\paragraph*{\bf 5. Solve the remaining equations for the periodic protocol.} Equations in (\ref{FIm}) are identical to  Eqs.~(\ref{Icons0})-(\ref{del-theta}) with the parameter $\mu$ replaced with $\cos \vartheta$. If we assume that the system initially performs oscillations along the vector $\hat{\theta}$, then after the full rotation of the structure, the oscillations along $\hat{\varphi}$ and $\hat{\vartheta}$ will be described by the actions from Eq.~(\ref{delI1-fin}), in which $\mu$ is replaced with $\cos \vartheta$. The Foucault rotation angle, $\phi_F$, is then defined analogously to Eq.~(\ref{xx-meas}) at $m_1=m_2=m$:
$$
\tan \phi_F=\frac{Y}{X}=-\tan (2\pi \cos \vartheta).
$$
Thus, by completion of the rotation of ${\hat{r}}(t)$ counterclockwise around the fixed $z$-axis, we achieve  
\begin{equation}
\phi_F=-2\pi \cos\vartheta.
\label{rot-angle}
\end{equation}
  In the fixed frame (relative to distant stars),  an observer would see that, during the same time, the plane of motion with vectors $\hat{\varphi}$ and $\hat{\vartheta}$ rotates around $\hat{z}$ axis by an additional angle $2\pi$.  Therefore, the physical trajectory in the fixed frame experiences the rotation by an angle $2\pi(1-\cos \vartheta)=\Omega_{sa}$. 

  Up to the minus sign, this is the famous expression for the rotation of the Foucault pendulum in terms of the Earth co-latitude $\vartheta$. The sign difference appears because, conventionally, the natural clockwise rotation of the Earth is countered as positive, wheres we assumed a more common convention for laboratory experiments that the rotation angle is counted positive for the counterclockwise motion of the rotated frame.  

Our derivation of this geometric angle is certainly not the shortest known one, but it proves, in addition, the conservation of the area inside the in-plane trajectory and reveals the relation of Foucault pendulum rotation to a broader class of mechanical geometric phase effects that are generally not described by conventional Hannay angles.

\section{Discussion}
We conclude with a conceptual summary of our approach. Consider any integrable mechanical system with many degrees of freedom and, initially, a time-independent Hamiltonian, $H$.  Integrability means that, in certain variables, $H$ can be expressed as the sum in Eq.~(\ref{Hsep}).  When switching to the action-angle variables $(I_k,\theta_k)$, this Hamiltonian takes the form
\begin{equation}
H=\sum_{k=1}^N H_k(I_k).
\end{equation}

 This system is characterized by a set of frequencies (\ref{freq1}). 
 If all frequencies are incommensurate, then all actions $I_k$ are separately conserved during the adiabatic evolution. Namely, the nonadiabatic contribution to the Hamiltonian in this case couples the variables that oscillate with incommensurate frequencies. In the equations of motion for $I_k$, this  leads to terms that contain products of quickly oscillating functions, such as $\propto \sin \omega_{k}t \sin \omega_{k'}t$, which vary quickly with time around zero mean value. As discussed in Ref.~\cite{landau}, the corrections to $I_k$ introduced by such terms in the quasi-adiabatic regime are suppressed exponentially; therefore,  in the adiabatic limit, they are neglected. The only geometric effects that are expected then are the standard Hannay angles for the quickly changing angle variables.

 Let us now consider a subsystem with variables having commensurate  frequencies. We choose their indices to be $s=1,\ldots,M$. For slowly time-dependent parameters, the nonadiabatic contribution to the Hamiltonian may lead to the terms in equations for $I_s$ that do not average to zero. For example, in the present article, we found that for degenerate frequencies,  $\omega_k=\omega_{k'}$, the physical rotation of a mechanical system generally leads to the terms that behave as $\propto \sin \omega_kt \sin \omega_{k'}t=\sin^2 \omega_kt$, which time-averages to $1/2$ rather than to zero. Hence, such terms cannot be disregarded.  In the case of nondegenerate but commensurate frequencies, e.g., $\omega_k=2\omega_{k'}$, a simple physical rotation of our system would be insufficient to induce this effect, but we cannot exclude the existence of less trivial integrable systems, in which the nonadiabatic contribution to the Hamiltonian leads to tri-linear coupling terms, which in turn induce relevant  contributions to the evolution of $I_k$ and $I_{k'}$, such as  $\propto \cos \omega_kt \sin^2 \omega_{k'}t$.  For  $\omega_k=2\omega_{k'}$, this would also not average to zero.

 To address such situations generally, we should  consider the effective Hamiltonian only for the commensurate frequency variables, 
 \begin{equation}
H_{\rm eff}=\sum_{s=1}^M H_s(I_s),    
\end{equation}
where
\begin{equation}  
n_s\omega_s=n_{s'}\omega_{s'},\quad \forall s,s'=1,\ldots,M,
\label{comm-cond}
\end{equation}
and where $n_{s}$ and $n_{s'}$ are integers. 
At such conditions, the Hamiltonian depends only on the combination
\begin{equation}
 I=\sum_{s=1}^M n_sI_s,
 \label{Icons}
\end{equation}
that is, 
\begin{equation}
H_{\rm eff}=H_{\rm eff}(I).
\label{heff}
\end{equation}
This follows from the definitions in Eqs.~(\ref{freq1}), (\ref{Icons}), and relations in Eq.~(\ref{comm-cond}). There is a corresponding angle variable $\theta$  that changes rapidly with frequency, 
$$
\omega = \frac{\partial H_{\rm eff}}{\partial I}.
$$
In addition to $I$, for fixed parameters, any other combination of $I_s$ is an independent invariant of motion. Let us choose $M-1$ such different invariants and denote them as ${\cal I}_s$, where $s=1,\ldots,M-1$. We denote the full set of them as $\{{\cal I}\}$. There are also $M-1$ independent invariants that can be constructed from the angle variables because 
$$
\theta_s n_{s'}-\theta_{s'}n_s
$$
is conserved by the equations of motion with the Hamiltonian (\ref{heff}). Let us denote such $M-1$ combinations by $\vartheta_s$, $s=1,\ldots,M-1$. We refer to their full set as $\{\vartheta\}$.
 
Assume now that the original Hamiltonian is slowly time-dependent, so that, if time is treated as a parameter, there is  a certain explicitly time-dependent canonical transformation separating the variables, while the considered $M$ frequencies remain commensurate. Let $\lambda_i(t)$, $i=1,\ldots,n$ be the set of all $n$ time-dependent control parameters, and  $\{ \lambda \}$ be  their full set.

When switching to the canonical variables $I_s$ and $\theta_s$, the effective Hamiltonian acquires  non-adiabatic corrections: 
\begin{equation}
    H_{\rm eff}'=H_{\rm eff} +\sum_{i=1}^{n}\Lambda_i (\{I\},\{\theta\})\dot{\lambda}_i.
\end{equation}
In the $2M$ corresponding equations of motion
$$
\dot{I}_s = -\frac{\partial   H_{\rm eff}'}{\partial \theta_s}, \quad \dot{\theta}_s =\frac{\partial H_{\rm eff}'}{\partial I_k}, \quad s=1,\ldots, M, 
$$
we switch to the variables $I$, $\theta$, $\{{\cal I}\}$, and  $\{\vartheta \}$. Then we average the result over the period of fast oscillations in the factors that depend on $\theta(t) \sim \omega t$.

This yields, for such averaged variables, that $\dot{I}=0$, while the $2(M-1)$ ``slow" variables, $\{{\cal I}\}$ and $\{{\vartheta}\}$, satisfy a set of coupled equations of the form 
\begin{eqnarray}
\dot{\cal I}_s &=&\sum_{i=1}^n F_{si}(\{{\cal I}\},\{\vartheta\};\{\lambda\}) \dot{\lambda}_i ,\\
\dot{\vartheta}_s &=&\sum_{i=1}^n G_{si}(\{{\cal I}\},\{\vartheta\};\{\lambda\}) \dot{\lambda}_i, \quad s=1,\ldots,M-1,
\end{eqnarray}
where the functions $F_{si}$ and $G_{si}$  can depend on all control parameters. Generally, such equations  do not decouple from one another. After the cyclic evolution of the control parameters, the final changes of $\{ I\}$ and $\{\vartheta \}$ do not depend on time but rather depend on the choice of the control protocol.  This generally leads to a non-Abelian geometric phase effect. For special choices of the control protocol, additional adiabatic invariants may exist, which reduce the complexity of this geometric phase, as in the examples from our article.  

Finally, we mention that the ability of both quantum and classical systems to non-trivially change their states upon quasi-adiabatic changes of external conditions is of considerable practical interest for building  more sensitive gyroscopes -- the devices measuring rotation angles.  They are used, for example, to control orientations of airplanes and satellites. 
%Such applications may motivate students to learn about the unusual properties of classical integrable systems that we have described in this work.

\begin{acknowledgements}
We thank Andrei Piryatinski for helpful discussions. This work was supported primarily by the U.S. Department of Energy, Office of Science, Office of Advanced Scientific Computing Research, through the Quantum Internet to
Accelerate Scientific Discovery Program, and in part by the
U.S. Department of Energy, Office of Science, Basic Energy
Sciences, under Award Number DE-SC0022134.
F.S. acknowledges support from the Los Alamos National Laboratory LDRD program under project number 20230049DR
and the Center for Nonlinear Studies under project number
20250614CR-NLS.
\end{acknowledgements}


\begin{thebibliography}{100}
\bibitem{note} Here, we do not consider situations that lead to breakdown of the adiabatic approximation, e.g., near crossing a separatrix.
\end{thebibliography}
\end{document}